\documentclass[conference,a4paper,flushend]{iaria} % arxiv: DIN A4 (a4paper)
\pdfoutput=1 % pdflatex hint for arxiv.org (within first 5 lines)

\usepackage[babel=true,english=american]{csquotes}
\usepackage[USenglish]{babel}
\usepackage[alwaysadjust]{paralist}
\usepackage[
  babel=true, % Enable language-specific kerning
  expansion=alltext,
  protrusion=alltext-nott, % Ensure no changes at the edge of the listing.
  nopatch=eqnum, % fix unable to apply patch eqnum
  final % Always enable microtype, even if in draft mode.
]{microtype}
\usepackage{upquote}
\usepackage[zerostyle=b,scaled=.75]{newtxtt}
\DisableLigatures{encoding = T1, family = tt* }
\usepackage[shortcuts]{extdash} % Use \-/ for a breakable dash that does not prevent the remainer of the word to be hyphenated
\usepackage{relsize} % allows for \textsmaller{..}
\newcommand{\eg}{e.g.,\ }

\usepackage{booktabs}
\usepackage{amsmath,amssymb,amsfonts} % must be loaded before cleveref
\usepackage[capitalise,nameinlink,noabbrev]{cleveref}
\usepackage{stfloats} % floats in a twocolumn document with [t] and/or [b]
\fnbelowfloat % Put footnotes below floats
\usepackage[bb=boondox,bbscaled=.95,cal=boondoxo]{mathalpha}
\definecolor{OliveGreen}{HTML}{3C8031}
\definecolor{Sepia}{HTML}{671800}
\definecolor{LightGray}{rgb}{0.97,0.97,0.97}

\usepackage{listings} % BEWARE: has known limitations with UTF-8 characters!

\lstdefinelanguage{SPARQL}{
  morekeywords={
    PREFIX, SELECT, CONSTRUCT, WHERE, FILTER, OPTIONAL, UNION,
    BIND, GROUP, BY, HAVING, AS, a, VALUES, ORDER, LIMIT, OFFSET, DISTINCT
  },
  sensitive=true,   % language is case sensitive
  comment=[l][\itshape]{\#},
  alsoletter={?,$}, % treat ? and $ as valid parts of identifiers (expects raw characters; do NOT escape!)
  morestring=[b]",  % double quotes (") define string literals, highlighted as strings
  morecomment=[l][\color{gray}]{\#},       % comments
  morecomment=[n][\color{blue}]{<http}{>}, % uris
  keywordsprefix=?,
  classoffset=0,
  keywordstyle=\color{Sepia},
  morekeywords={},
  deletekeywords={
    SELECT,CONSTRUCT,DESCRIBE,ASK,WHERE,FROM,NAMED,BASE,OPTIONAL,
    FILTER,GRAPH,LIMIT,OFFSET,SERVICE,UNION,EXISTS,NOT,BINDINGS,MINUS,a
  },
  classoffset=1,
  keywordstyle=\color{MidnightBlue},
  morekeywords={
    SELECT,CONSTRUCT,DESCRIBE,ASK,WHERE,FROM,NAMED,PREFIX,BASE,OPTIONAL,
    FILTER,GRAPH,LIMIT,OFFSET,SERVICE,UNION,EXISTS,NOT,BINDINGS,MINUS,a
  }
}
\DeclareCaptionType{listing}[lstlisting][List of Listings]
\vbadness=\maxdimen
\usepackage{algorithm}
\usepackage{algpseudocode}

\usepackage{ifpdf}
\ifpdf
\pdftrue
\fi

\makeatletter
\let\blx@rerun@biber\relax
\makeatother

\makeatletter
\def\ps@IEEEtitlepagestyle{
    \def\@oddfoot{\mycopyrightnotice}
    \def\@evenfoot{}
}
\def\mycopyrightnotice{
    {\footnotesize
        \begin{minipage}{0.8\textwidth}
            \centering
            Please cite as: \fullcite{selfref}.
        \end{minipage}
    }
}
\makeatother
\usepackage[shortcuts]{extdash} % Use \-/ for a breakable dash that does not prevent the remainer of the word to be hyphenated

\DeclareBibliographyCategory{selfref}
\addtocategory{selfref}{selfref}
\usepackage[nolist]{acronym} % ,smaller

\begin{acronym}
\acro{abi}[ABI]{Application Binary Interface}
\acro{api}[API]{Application Programming Interface}
\acro{evm}[EVM]{Ethereum Virtual Machine}
\acro{utxo}[UTXO]{Unspent Transaction Output}
\acro{kg}[KG]{Knowledge Graph}
\acro{llm}[LLM]{Large Language Model}
\acro{rdf}[RDF]{Resource Description Framework}
\acro{owl}[OWL]{Web Ontology Language}
\acro{defi}[DeFi]{Decentralized Finance}
\acro{ba}[BA]{blockchain address}
\acrodefplural{ba}[BAs]{blockchain addresses}
\acro{rwe}[RWE]{real-world entity}
\acrodefplural{rwe}[RWEs]{real-world entities}
\acro{ft}[FT]{Fungible Token}
\acro{nft}[NFT]{Non-Fungible Token}
\acro{eoa}[EOA]{Externally Owned Account}
\acro{ner}[NER]{Named Entity Recognition}
\acro{eth}[ETH]{Ether}
\end{acronym}

\usepackage{booktabs}

\title{The Kosmosis Approach to Crypto Rug Pull Detection}

\author{
  \IEEEauthorblockN{
    ~\\[-0.3ex]
    Philipp Stangl\IEEEauthorrefmark{1}\,\orcidlink{0009-0007-4179-2365} and
    Christoph P.\ Neumann\IEEEauthorrefmark{2}\,\orcidlink{0000-0002-5936-631X}
    \\[0.3ex]~
  }
  \IEEEauthorblockA{\IEEEauthorrefmark{1}%
    Department of Computer Science\\
    Friedrich-Alexander-Universität Erlangen-Nürnberg, Erlangen, Germany\\
    e-mail: {\tt philipp.stangl@fau.de}
  }
  \IEEEauthorblockA{\IEEEauthorrefmark{2}%
    Department of Electrical Engineering, Media and Computer Science\\
    Ostbayerische Technische Hochschule Amberg-Weiden, Amberg, Germany\\
    e-mail: {\tt c.neumann@oth-aw.de}
} }

\begin{document}

\maketitle

\begin{abstract}
Crypto rug pulls have become a major threat to the integrity of blockchain ecosystems, with illicit activities surging and resulting in significant financial losses. 
Existing approaches to detect crypto asset fraud are based on the analysis of transaction graphs within blockchain networks.
While effective for identifying transaction patterns indicative of fraud, existing approaches do not capture the semantics of transactions and are constrained to blockchain data.
Consequently, preventive methods based on transaction graphs are inherently limited.
In response to these limitations, we propose the Kosmosis approach, which aims to incrementally construct a knowledge graph as new blockchain and social media data become available.
During construction, it aims to extract the semantics of transactions and connects blockchain addresses to their real-world entities by fusing blockchain and social media data in a knowledge graph.
This enables novel preventive methods against rug pulls as a form of crypto asset fraud.
To demonstrate the effectiveness and practical applicability of the Kosmosis approach, we examine a series of real-world rug pulls.
Through this case, we illustrate how Kosmosis can aid in identifying such fraudulent activities by leveraging the insights from the constructed knowledge graph.
\end{abstract}

% A list of IEEE Computer Society appoved keywords can be obtained at
% http://www.computer.org/mc/keywords/keywords.htm

%\blfootnote{Note: This article is a revised and extended version of \cite{StNe25kosmosisGraphConstruction} and \cite{StNe25kosmosisUseCase}}

\section{Introduction}
This article is a revised and extended version of \cite{StNe25kosmosisGraphConstruction} and \cite{StNe25kosmosisUseCase}.
Crypto assets are digital assets that use distributed ledger technology, such as blockchain, to prove ownership and maintain a decentralized and public ledger of all transactions.
%There are distinct types of assets, each with unique characteristics and use cases. 
Cryptocurrencies, like Bitcoin~\cite{nakamoto2008}, function as digital currencies and are used for storing or transferring monetary value. 
\acp{ft}, another type of crypto asset, are interchangeable units representing various utilities or assets within a blockchain ecosystem. 
%These tokens often play a vital role in \ac{defi} protocols and can represent anything from voting rights to a currency within a project ecosystem. 
Lastly, \acp{nft} are unique digital assets that prove ownership and authenticity of digital or real-world assets \cite{alizadeh2023}. 
%Unlike cryptocurrencies and \acp{ft}, each \ac{nft} has a distinct value and cannot be exchanged on a one-to-one basis with other tokens.
In the rapidly evolving landscape of crypto assets, the incidence of illicit activities has surged. Chainalysis, a leading blockchain analytics firm, reported that illicit transaction volume rose for the second consecutive year in 2022, reaching an all-time high of \$20.6 billion in illicit activity \cite{chainalysis2023}.
Since the rise of \ac{defi} in 2020, followed by \acp{nft} in 2021, rug pulls have become a major fraud scheme in terms of amount stolen and frequency \cite{sharma2023}.
Thus, rug pulls pose a significant risk to investors and undermine the integrity of the crypto asset sector. 

The predominant approach for identifying patterns indicative of fraudulent activity is the transaction graph analysis within blockchain networks
%\cite{khan2022, beres2021}\cite[pp. 21--24]{huang2022}. 
\cite{khan2022, beres2021, huang2022}.
However, this approach presents two key challenges. 
Firstly, the transacting parties are pseudonymous and only their \aclp{ba} are publicly known.
This means that, although the transactions of a specific address can be tracked, linking that address to a \acl{rwe} can be challenging since this approach is limited to information or patterns observable in blockchain data.
Secondly, this approach is only concerned with the following aspects of a transaction: 1) The transferred asset, 2) the quantity, and 3) the sender and receiver.
However, the semantics of a transaction, such as what happened in a transaction that caused the assets to get transferred, is not covered. 
Thereby limiting the depth of analysis that can be conducted on crypto asset movements. 

\Acp{kg} \cite{hogan2022} are increasingly recognized as a powerful means to integrate fragmented knowledge from heterogeneous data sources into a graph of data to facilitate
semantic querying
(\eg \cite{ModA-TR-2023WS-SWT-TeamRot-SGDb,ModA-TR-2022WS-SWT-TeamGruen-Graphvio})
and reasoning
(\eg \cite{NeFL10oxdbs}).
A \ac{kg} provides a holistic view for identifying patterns and hidden connections indicative of fraudulent activities in a highly connected dataset \cite{zhu2021}.
The \ac{kg} consists of semantically described entities, each with a unique identifier, and relations among those entities using an ontological representation \cite{feilmayr2016, hofer2023}.
Their open world assumption allows for the continual integration of new data.
By leveraging these capabilities, \acp{kg} can enhance crypto asset fraud analysis and aid in predicting future fraudulent activities.

The remainder of this article is organized as follows. 
We first outline the Kosmosis objectives in Section~\ref{s:objectives}.
Next, we provide a background on the Ethereum blockchain and graph-based blockchain data mining methods in Section~\ref{s:background}. 
Subsequently, in Section~\ref{s:kosmosis} we propose Kosmosis, our incremental \ac{kg} construction pipeline.
In Section~\ref{s:rug_pulls} we provide essential background on rug pulls before we demonstrate the effectiveness and practical applicability of the Kosmosis approach for the use case of \ac{nft} rug pull detection in Section~\ref{s:use_case}.
Finally, we outline future work in Section~\ref{s:future_work} and conclude the article with a discussion of our findings in Section~\ref{s:conclusion}.

\begin{figure*}[b!]
\centerline{\includegraphics[width=.75\textwidth]{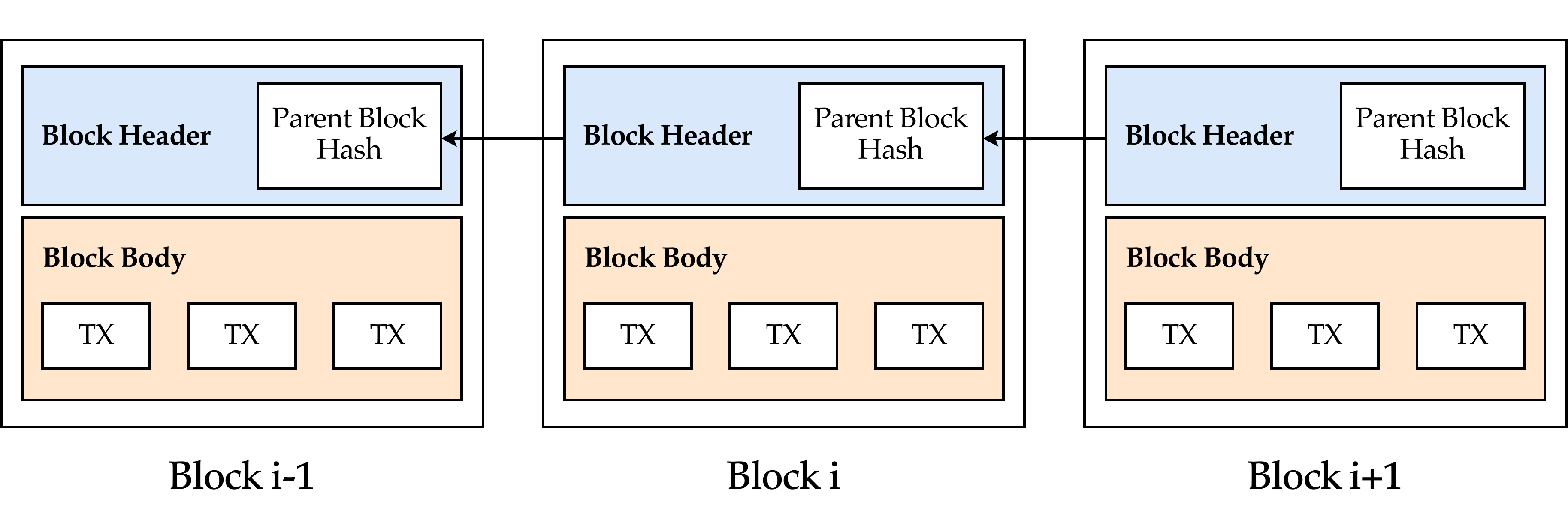}}
\caption{Schematic representation of the blockchain data structure. Adapted from \textcite{zheng2017}.}
\label{fig:datastructure}
\end{figure*}

\section{Overarching Objectives of Kosmosis} \label{s:objectives}

This section outlines the objectives of Kosmosis, beginning with the primary objective that investigates the potential of a \ac{kg} in identifying and alerting users before they interact with projects linked to known scammers, addressing a critical need for security and trust in blockchain ecosystems. 
Following that, we explore the technical implications.

\textbf{Objective 1: How can the \ac{kg} identify and aid in alerting users before interacting with a rug pull project?}

With the rise in illicit activities in the crypto asset market, especially rug pulls, there is a pressing need for effective means to detect and prevent fraudulent activities. Kosmosis aims to integrate fragmented knowledge from blockchains like Ethereum, social media like $\mathbb{X}$, and potentially other knowledge graphs into one unified \ac{kg}, enabling semantic querying and reasoning over a graph of entities and the relationships among them. 
The \ac{kg} could serve as a knowledge base for a real-time alerting system, warning users of potential risks associated with certain projects or individuals. 

\textbf{Objective 2: How to incrementally construct the \ac{kg} from heterogeneous data sources?}

It is imperative to establish a pipeline capable of integrating updates into the \ac{kg} in both batch- and streaming-like manner.
Thereby, maintaining high data freshness by ensuring that the \ac{kg} consistently reflects the most up-to-date information from the blockchain and other sources.
This approach should not entail a complete reconstruction of the \ac{kg}, but rather concentrate on integrating new information, avoiding the reprocessing of data that is already incorporated. 

\textbf{Objective 3: How to extract the semantics of blockchain transactions?}

Transaction graphs commonly only display transactions with asset transfers and provide answers to questions such as ``what'' assets were transferred, and ``where'' were they transferred to.
Understanding transactions semantically is vital in uncovering sophisticated fraudulent schemes that might otherwise go unnoticed.
Kosmosis addresses this gap by extracting the semantics of transactions, providing answers to ``why'' and ``how'' assets were transferred in a transaction.
This extraction of semantic information is primarily achieved through decoding the input data of a transaction using the \ac{abi} of smart contracts a transaction interacts with.

\section{Background} \label{s:background}

In this section we provide background on blockchain technology, specifically the Ethereum blockchain, in Section~\ref{s:blockchain_technology}. 
Subsequently, we outline related graph-based blockchain data mining methods in Section~\ref{s:graph_blockchain_data_mining}.
On the social media aspects of Kosmosis, our prior work includes correlating Reddit data with traditional stock market trends \cite{ModA-TR-2022SS-BDCC-TeamRot-Reddiment} and analyzing Twitter/$\mathbb{X}$ data using SPARQL \cite{ModA-TR-2022SS-BDCC-TeamWeiss-TwitterDash}.

\subsection{The Ethereum Blockchain} \label{s:blockchain_technology}
Blockchain technology is based on the principles of immutability, decentralization, transparency, and cryptographic security and has seen various applications in recent years.
For instance, in the financial sector (e.\,g., \cite{nakamoto2008, wood2024}), or supply chain management (e.\,g., using a single blockchain \cite{wang2019},
or using multiple, interoperable blockchains
\cite{StNe23foodfresh,stangl_philipp_2022bt_en}).
Smart contract platforms represent a subset of blockchains that enable the development of decentralized applications through smart contracts. 
This section outlines the key concepts of Ethereum, as an example for smart contract platforms, that are essential for the following sections of this work, such as smart contracts, their execution environment, and account-based accounting.

\subsubsection{Blockchain Data Structure}
A blockchain is a data structure whose elements called blocks are linked together to form a chain of blocks \cite{zheng2017}, depicted in Figure~\ref{fig:datastructure}.
Each block comprises two parts: a body and a header. 
The body of the block contains a set of transactions.
A transaction typically involves the transfer of assets between a sender and a receiver. 
These participants are represented by addresses, which are unique alphanumeric strings that clearly specify the origin and destination of each transaction.
Further, the block body is used to generate a unique identifier called the block hash.  
The block header contains a reference to the unique identifier of its immediate predecessor, known as the parent block. 

\begin{figure*}[b!]
\centerline{\includegraphics[scale=0.82]{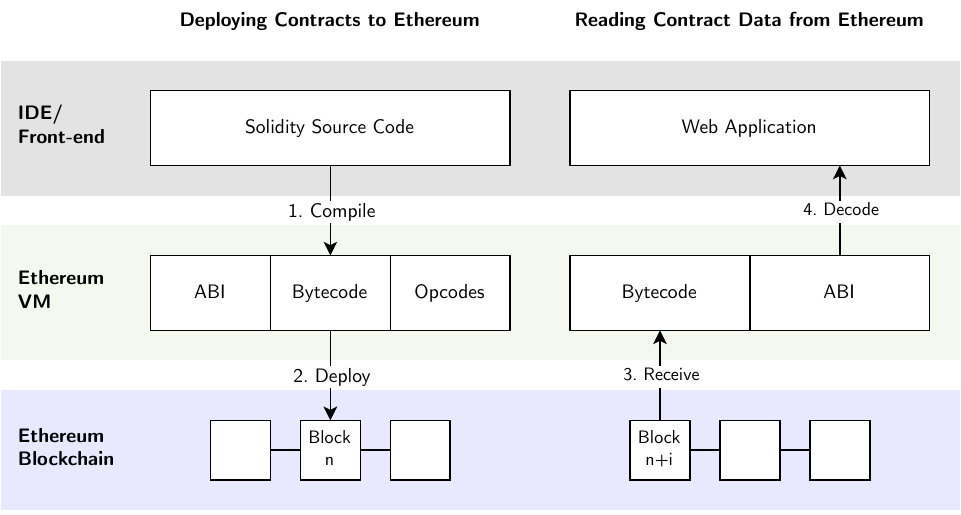}}
\caption{Schematic representation of deploying and reading from smart contracts. Adapted from \textcite{eiki2023}.}
\label{fig:evm}
\end{figure*}

\subsubsection{Smart Contracts}
Through smart contracts, which are executable source codes that enforce the terms and conditions of particular agreements, a smart contract platform like Ethereum facilitates the development of decentralized applications \cite{marin2023}. 
Once deployed on the blockchain, the smart contract is assigned an address where the code resides and cannot be altered or tampered with.
%Standards for the Ethereum platform are described in \acp{erc}.
By writing custom smart contracts, developers can create and manage tokens that adhere to the standards ERC-20 for \acp{ft} \cite{vogelsteller2015}
or ERC-721 for \acp{nft} \cite{entriken2018}.
An \ac{abi} specifies the functions and data structures exposed by a smart contract, allowing external applications to understand the capabilities of the contract. 
Further, an \ac{abi} defines a format for encoding and decoding data that is passed between smart contracts and external applications. 
This ensures a consistent and standardized way to exchange information.

The Ethereum blockchain manages \ac{eth} as the native cryptocurrency of the platform.
It operates with the \ac{evm} as a fundamental building block, serving as the execution environment for smart contract code. 
Smart contracts, primarily written in a high-level language such as Solidity, undergo compilation into \ac{evm} bytecode. 
This bytecode is the executable format used by the \ac{evm} to enact smart contract functions. 
To interact with this bytecode, a contract \ac{abi} is utilized, which acts as a bridge between the high-level language and the low-level bytecode. 
In this context, an \ac{evm} disassembler plays a crucial role; it reverses the bytecode back into a more readable format, aiding developers in understanding and analyzing the code deployed on the Ethereum blockchain. 
Figure~\ref{fig:evm} shows the processes involved in deploying smart contracts to the Ethereum blockchain and reading contract data from it, including compilation and deployment steps, and the interaction between a web application and the Ethereum blockchain.
The left side shows the compilation and deployment of a smart contract, and the right side depicts an interaction with the contract (\eg from a web application).

\subsubsection{Externally Owned Account}
Unlike smart contracts, \acp{eoa} are controlled by \aclp{rwe} through private keys, enabling them to initiate transactions, such as transferring crypto assets or executing functions of a smart contract. 
When an \ac{eoa} sends a transaction to a smart contract, it triggers the code of the contract to execute according to its predefined rules. 

\subsubsection{Account-based Accounting} \label{s:blockchain_accounting}
For the record-keeping of transactions, blockchains utilize an accounting model.
Compared to other blockchains, such as the equally well-known Bitcoin \cite{nakamoto2008} blockchain that uses the \ac{utxo} model, or its successor the extended \ac{utxo} \cite{chakravarty2020} utilized by the Cardano \cite{hoskinson2017} blockchain, Ethereum \cite{wood2024} employs the account-based accounting model.
The account-based model can be best understood through the analogy of a bank account. 
This approach mirrors how a banking account operates.
Like a bank account that tracks the inflow and outflow of funds, thereby reflecting the current balance, the account-based model in Ethereum maintains a state that records the balance of Ether. 
Thus, it is inherently stateful. 
Each transaction results in a direct adjustment to this balance, akin to a deposit or withdrawal in a bank account. 
This model's stateful nature ensures that at any given moment, the system can accurately reflect the total amount of Ether held in each account, offering an up-to-date view of account balances within Ethereum. 

\subsubsection{Token Minting}

The process of creating new tokens is called token minting.
\acp{ft} are typically minted by the creator either at the inception of the project or progressively over time. 
This process is often governed by predefined rules or algorithms embedded within the smart contracts of the project.

In contrast, \ac{nft} minting involves other individuals besides the token creator, commonly termed as token minters. 
They engage by invoking a specific function within a smart contract, in the ERC-721 token standard, called mint. 
This action results in an increase in the supply of the \acp{nft} and simultaneously assigns these minted tokens to the \acl{ba} of the minter.
The mechanism of minting \acp{nft} often involves utilizing a dedicated minting website. Here, prospective minters or investors are required to invest a predetermined amount, as set by the creator, to initiate the minting process. This investment grants them the ability to mint one or multiple \acp{nft}, depending on the terms set forth in the smart contract. 
%This process not only facilitates the creation of new \acp{nft} but also serves as a means of transferring ownership directly from the creator to the \ac{nft} minter.

\subsection{Rug Pull Detection Methods} \label{s:graph_blockchain_data_mining}

Two primary methods have been employed in the past to detect rug pulls: smart contract code analysis and graph-based methods.
Smart contract code analysis involves a thorough examination of the contract’s code to extract and analyze the semantic behavior of transactions. For instance, \cite{zhou2024} utilizes smart contract code analysis to reveal potential vulnerabilities and fraudulent patterns within the contracts. By dissecting the code, their proposed method, dubbed \enquote{Tokeer}, can identify suspicious patterns and functions that might indicate a predisposition to rug pull scams.
Another prominent line of research in smart contract analysis leverages machine learning–based techniques. 
\citeauthor{mazorra2022rug} \cite{mazorra2022rug}, for example, employed the XGBoost algorithm as a primary classifier to predict the probability that a liquidity pool will evolve into a rug pull or scam, achieving an accuracy of up to 99.36\% using features derived from token propagation patterns and smart contract heuristics. 
Their dataset and experimental design are restricted to fungible tokens launched exclusively on Uniswap (Ethereum), and the authors argue that directly transferring these learned models to other blockchains is unlikely to yield results of comparable quality. 
%Moreover, because the computation of the clustering coefficient is highly time-consuming, the authors propose using other graph analysis methods to obtain a more efficient detection algorithm.
Graph theories and graph-based data mining methods, are applicable for discovering information in blockchain network graphs, because blockchain transactions can be inherently structured into graphs \cite{huang2022}.
\citeauthor{elmougy2023} \cite{elmougy2023} identified three types of graphs, applicable to any blockchain network: \emph{money flow transaction graphs} visualize the asset flow over time, \emph{address-transaction graphs} showing flow of an asset across transactions and addresses, and \emph{user entity graphs} that clusters the graph for potential linking of addresses controlled by the same user, to deanonymize their identity and purpose.
To detect rug pulls with high accuracy, graph-based approaches use network embedding methods to automatically extract features from the blockchain network (e.\,g., \cite{chenliang2020}) using a \textit{graph convolutional network}.

\section{The Kosmosis Approach to Incremental Knowledge Graph Construction} \label{s:kosmosis}

\begin{figure*}[tb]
\centerline{\includegraphics[width=\textwidth]{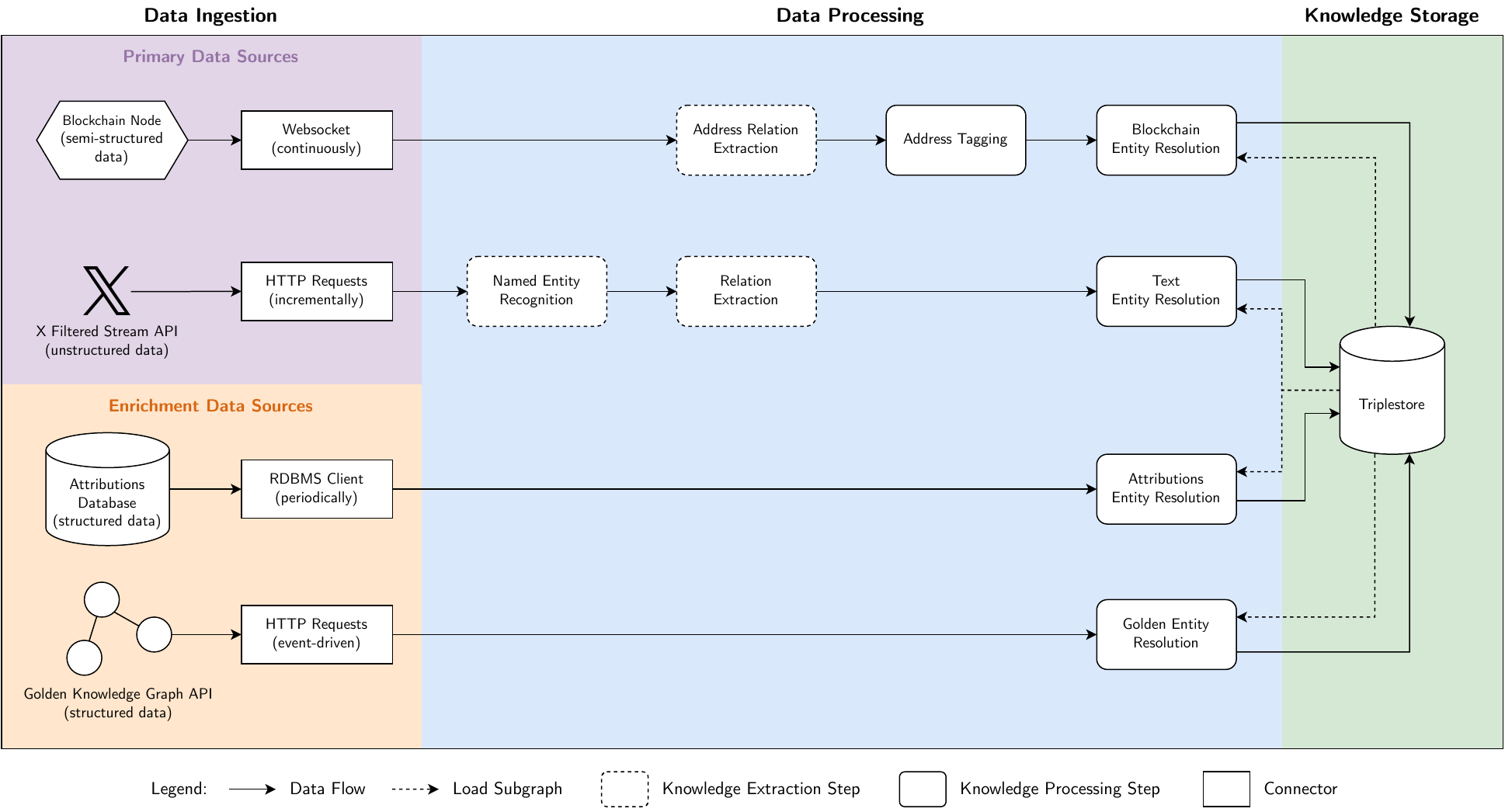}}
\caption{A high level overview of the Kosmosis pipeline.}
\label{fig:kosmosis_pipeline}
\end{figure*}

To incrementally construct a \ac{kg} that integrates data in a continuous and periodic way, we propose a multi-stage pipeline, as illustrated in Figure \ref{fig:kosmosis_pipeline}.
It originated from a master's thesis \cite{stangl_philipp_2024mt} and consists of three stages: Data ingestion, data processing, and knowledge storage.
We use italics to emphasize on conceptual aspects and typewriter text for technical operations.

The initial stage, data ingestion, captures the raw data from the primary data sources (blockchain and social media) as well as enrichment data sources (e.g., another knowledge base). 
This phase is characterized by its versatility in the frequency of data acquisition: it can be 1) \emph{continuous}, to capture real-time updates from sources such as blockchain nodes, 2) \emph{incremental} for new posts via the $\mathbb{X}$ Streaming \ac{api}, 3) \emph{periodic}, to capture new entries in structured data sources like relational databases at regular intervals, or 4) \emph{event-based}, responding to events that are emitted upon new entity additions to the \ac{kg}.

Following the ingestion stage, the data processing stage is initiated, which is partitioned into distinct workflows tailored to handle each type of ingested data. 
This segmentation allows for specialized processing depending on the structure of the raw data. 
For instance, for text data sources, natural language processing techniques, such as \acl{ner} \cite{li2020survey}, can be used to ensure that the data is accurately interpreted, and contextual relationships are discerned.

In the third and final stage, the refined data is loaded into the knowledge storage, where it is systematically organized within a triplestore, a type of database optimized for storing and retrieving data in \ac{rdf} format.
The triplestore can then be used for semantic querying capabilities to extract actionable insights from the \ac{kg} for downstream processes. 
For the \ac{kg}, we use the EthOn \cite{ethon2023} ontology that formalizes the concepts and relations within the domain of the Ethereum network and blockchain.
EthOn is written in \ac{rdf} and \ac{owl}.

\subsection{Blockchain Data Processing}

The blockchain data processing workflow continuously ingests new transactions from the blockchain via websocket connections. Websockets enable open, interactive communication sessions between a client and a server, facilitating real-time data transfer without the need for repeated polling. Upon receiving these transactions, the workflow processes and integrates them into the \ac{kg} by first extracting the address relationship, followed by tagging the addresses, and finally fusing the addresses with the entities of the \ac{kg}.

\subsubsection{Address Relation Extraction}

In order to provide answers to ``why'' and ``how'' assets were transferred in a transaction, Kosmosis implements a pipeline module titled \textit{Address Relation Extraction}.
The responsibility of this module is to extract the semantic information in a blockchain transaction through decoding the input data of a transaction using the \ac{abi} of the smart contract a \acl{ba} is interacting with.

First, the \ac{abi} is requested from Etherscan \cite{etherscan2023} and Sourcify \cite{sourcify2023} via their respective REST \acp{api}. 
If the \ac{abi} cannot be successfully fetched from one of the aforementioned sources, the module resorts to reconstructing the \ac{abi} from the smart contract byte code, which is available at any time since the bytecode is deployed on the blockchain.  
This operation enables the decoding of transactions and the interaction with smart contracts beyond their compiled state. 

The initial step involves the disassembly of the bytecode of the smart contract. This operation, referred to as \texttt{DISASM}, decomposes the bytecode into a series of readable opcodes and associated data. Disassemblers (e.g., \textit{pyevmasm} 
\cite{manzano2024}) facilitate this step by translating the bytecode back into a form that represents the original instructions and operations defined within the smart contract.

Following disassembly, the algorithm initializes by creating an empty array intended to store the \ac{abi} and defining lists of opcodes that either change the state or read from the state of the blockchain. These opcodes include \texttt{SSTORE}, \texttt{CREATE}, \texttt{CREATE2} for state-changing operations, and \texttt{SLOAD} for state-reading operations, reflecting the fundamental actions a smart contract on the \ac{evm} can perform \cite{wood2024}.

The core of the algorithm iterates over selector/offset pairs within the disassembled bytecode. Selectors serve as identifiers for functions in the \ac{evm}, facilitating the mapping to the corresponding functionality. If a given offset does not match any destination within the program's destinations, the iteration skips to the next pair, ensuring only valid functions are considered.

Upon finding a valid function destination, the algorithm retrieves the function definition and assigns tags based on its behavior. 
This tagging process involves analyzing the opcodes contained within the function and any related jump destinations. 
The purpose is to categorize functions according to how they alter the blockchain state, using a depth-first search algorithm to navigate through the function call graph.

An \texttt{AbiFunction} object is then created for each valid function, with its payable status determined inversely by the presence of a notPayable marker at the corresponding offset. The algorithm next assigns mutability attributes (nonpayable, payable, view, or pure) based on whether the function alters state, reads state, or neither. This classification is crucial for understanding how functions interact with the blockchain and their implications on transaction costs and permissions.

Finally, the algorithm decides on the inclusion of inputs and outputs in the function signature, informed by the presence of specific tags. For instance, tags indicating data retrieval or state mutation influence whether parameters are classified as inputs or outputs. This granular control ensures that the \ac{abi} accurately reflects the interface of the smart contract, allowing for effective transaction decoding.

Currently, the method for extracting semantic information from smart contract transactions relies partly on predefined heuristics, such as recognizing specific function names like ``mint.'' 
However, we acknowledge that scammers could circumvent these simplistic heuristics by obfuscating or renaming functions. 
Future improvements will incorporate advanced transaction pattern analysis rather than function naming alone, enhancing resilience against simple obfuscation techniques.

\subsubsection{Address Tagging}
Since the exact identity of a \acl{rwe} controlling a blockchain address is often unknown, it can still be categorized and tagged accordingly.
The \textit{address tagging module} tags the sender and receiver address based on their extracted relationship from the preceding address relation extraction module.
For instance, an \ac{eoa} deploying a smart contract is tagged as deployer in case of a contract creation transaction. 
Likewise, if an \ac{eoa} is sending Ether to an \ac{nft} contract $T$ via a contract function containing the word ``mint,'' the \ac{eoa} is tagged as is tagged as \ac{nft} minter of $T$.
Tags are subclasses of \acp{eoa} and contract accounts, extending the address concept of the EthOn ontology.

\subsubsection{Blockchain Entity Resolution}

The blockchain entity resolution module is responsible for resolving \aclp{ba} to either new entities or existing ones in the \ac{kg}, by using the extracted information from preceding steps.
It begins with mapping the result data from the preceding steps into the \ac{rdf} format, adhering to the ontology defined by the \ac{kg}. This ensures that the data is structured in a way that is compatible with the \ac{kg}'s existing schema.

Following the mapping to \ac{rdf}, the next phase involves fusing this \ac{rdf} data with the \ac{kg}. This is accomplished through a two-step process. Initially, a subgraph that is relevant to the processed data is loaded into the system. This step, commonly referred to as ``blocking,'' narrows down the scope of the resolution process to the most relevant segments of the \ac{kg}, thereby enhancing the entity resolution process.

Subsequently, the system proceeds to match the newly processed data with the corresponding entities within the \ac{kg}. This matching process is crucial for identifying where the new data fits within the existing structure and for ensuring that it is integrated in a meaningful way. In certain cases, the fusion process may also involve the clustering of entities. This is particularly relevant for blockchain data, where unique characteristics of the data can be leveraged to enhance the integration process.

For instance, when dealing with blockchains that utilize an account-based accounting model, address clustering heuristics can be employed to further refine the fusion process. One such heuristic is the deposit address reuse, as proposed by \citeauthor{victor2020} \cite{victor2020}. 
Kosmosis uses deposit address reuse for blockchain data from Ethereum to resolve entities more effectively.

\subsection{Text Processing}

The workflow starts with the input of unstructured data from the $\mathbb{X}$ Filtered Stream \ac{api} \cite{x2024}, which is incrementally streamed and parsed via a long-lived HTTP request into the pipeline. % for processing. 
The first step in processing this data is \acl{ner}, where the system identifies and classifies named entities present in the text into predefined categories, such as the names of persons, organizations, and locations.

The next step is relation extraction. This process involves identifying and extracting relationships between the named entities that were previously recognized. For instance, it could determine that a person named ``Alice'' works for a company named ``Acme.''

The final step in the text processing workflow is the entity resolution, achieved through blocking and matching.
For each new entity, the system identifies all other entities within the \ac{kg} that need to be considered for matching. 
Considering the growing size of the \ac{kg}, through the incremental updates, it is important to limit the matching process to as few candidates as possible \cite{hofer2023}. 
The method of limiting candidates is known as blocking, which confines the matching process to entities of the same or most similar entity type.

Following the blocking that serves as a preliminary filtering step, the matching is performed.
This involves a pairwise comparison of the new entities with those existing entities in the \ac{kg} identified during the blocking phase. 
Its objective is to identify all entities that are sufficiently similar and, therefore, potential candidates for matching. 
This pairwise comparison relies on a nuanced assessment of similarity that encompasses both the properties of the entities and their relational connections within the \ac{kg}. 
By evaluating both property values and the nature of relationships to other entities, the system determines the degree of similarity between entities.

\subsection{Enrichment Data Processing}

Enrichment data enhances the data obtained from primary data sources with supplementary context regarding \aclp{rwe}.
Attributions involve the mapping of blockchain addresses to their corresponding real-world entities. This task is largely dependent on data sourced from a network of experts, such as team members from blockchain projects.
The input data for the attribution process is typically not consistent in its timing, as it depends on when the experts provide updates or when new information becomes available. As a result, the enrichment data processing workflow is designed to operate at regular intervals, ensuring that the \ac{kg} is updated systematically and remains as up-to-date as possible.

To further enrich the \ac{kg}, data from external knowledge bases is integrated. In our case, we use the \emph{Golden Knowledge Graph} due to its concentrated information on tech startups and cryptocurrencies. This external graph offers a wealth of information about crypto projects, including details about their founders, team members, and project descriptions. Such depth of data provides a valuable context that can significantly improve the understanding of entities in the constructed \ac{kg}.

The workflow for integrating knowledge from an external \ac{kg} is event-driven, activated once the knowledge storage indicates the addition of new entities from the social media platform $\mathbb{X}$. Then, the workflow triggers a process to pull in additional background information from the Golden Enrichment \ac{api} \cite{golden2024}. It uses the $\mathbb{X}$ username that has been newly included in the \ac{kg} as unique identifier to fetch relevant data. 

\subsection{Quantity Structure of the Knowledge Graph Data}

In our prototype implementation, data was ingested at rates averaging 10-15 transactions per second (each averaging 5KB) from Ethereum blockchain nodes and roughly 200 tweets per minute (each averaging 2KB) from the $\mathbb{X}$ filtered stream API. 
This combined ingestion rate corresponds approximately between 3.4 to 4.9 MB per minute of raw data.
Our prototype runs on a standalone cloud server instance with 32 GiB RAM and 8 vCPUs (AWS EC2 m5.2xlarge) with a 512GB SSD, managing real-time data ingestion and processing workloads. 
The semantic enrichment introduces minimal latency (less than 5 seconds per transaction batch), thus allowing for near-real-time \ac{kg} updates.
The \ac{kg} constructed by Kosmosis accumulates triples at an approximate rate of 2.5 to 6 million triples per day, depending on transaction activity and the level of detail extracted from social media.

While the described hardware configuration proved adequate for prototype-level or small- to medium-scale deployments, a production implementation aimed at analyzing multiple blockchain networks or higher data volumes would necessitate scaling to multiple compute nodes, each handling dedicated tasks such as blockchain data ingestion.

\section{Rug Pulls and Serial Fraudsters} \label{s:rug_pulls}
A rug pull can be categorized as a scam, i.\,e., the victim authorizes the transaction.  
This type of scam is typically carried out in five stages, according to \cite{sharma2023}:
(1) Project creation with roadmap and total supply of tokens (optional), (2) pre-mint hype, (3) set token mint price, (4) token mint, accumulation of more capital and increase in popularity, and finally (5) the creators cash out, abandon the project, and leave the investors defrauded.
To attract users and investments for rug pulls, \citeauthor{sharma2023} \cite{sharma2023} suggest the involvement of individuals or groups that possess substantial technical skills and knowledge of blockchain technology and demonstrate a proficiency in marketing techniques.
This specific use case is particularly relevant given the findings in \cite{sharma2023} and \cite{mazorra2022rug}:
\citeauthor{mazorra2022rug}, who analyzed ERC-20 tokens listed on decentralized exchanges in their 2022 study, labeled 97.7\% out of 27,588 analyzed tokens as rug pulls \cite{mazorra2022rug}.
Likewise, \citeauthor{sharma2023} analyzed \acp{nft} and identified a cluster of 168 \acp{nft} associated with what they termed the ``Rug-Pull Mafia,'' a group of creators responsible for orchestrating multiple and repeated rug pulls \cite{sharma2023}. 
There is a growing trend in both the frequency and the financial impact of crypto rug pulls and scams \cite{comparitech2023}. 
Notably, the year 2021 marks a peak in the amount stolen, while 2022 shows a sharp rise in the frequency of these fraudulent activities and remains elevated since.

\begin{figure*}[b!]
\centerline{\includegraphics[width=\textwidth]{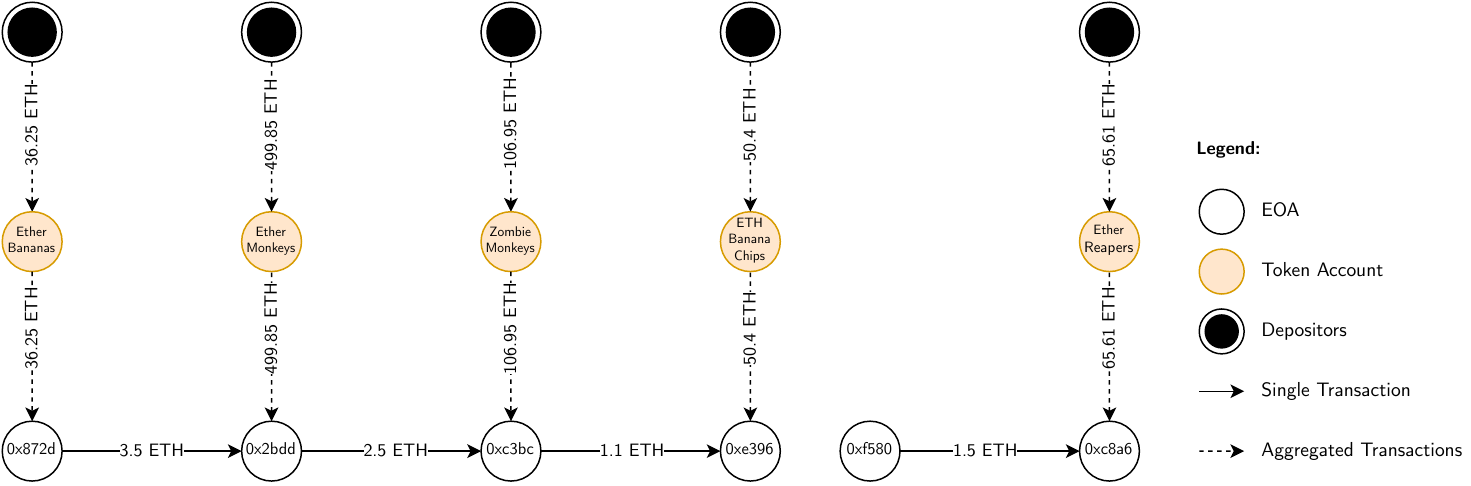}}
\caption{Simplified transaction graph of Homer\_eth's \ac{nft} rug pulls.}
\label{fig:use_case_tg}
\end{figure*}

\section{The Use Case of Rug Pull Prevention}
\label{s:use_case}

To illustrate the vision of Kosmosis-enabled rug pull prevention methods, this section introduces a hypothetical user story centered around a character we name Bob, a crypto market participant.
The story telling method of use case illustration was adopted from our previous work in \cite[pp.~207--209]{Neumann2013dissBook}.
The Kosmosis user story is designed to provide a relatable perspective on how individuals like Bob are affected by such fraudulent activities. 
The fictional story of Bob is grounded in a series of real-world rug pulls that took place in 2021. All rug pulls were carried out by the same fraudulent \ac{nft} creator and Twitter user known as Homer\_eth.
In Section \ref{sub:extended_user_story}, we outline how the series of rug pulls experienced by Bob might have unfolded differently had he been equipped with a Kosmosis-enabled fraud prevention mechanism at the time.

%\subsection{Rug Pull Case: NFT Creator Homer\_eth}
\subsection{Past User Story} \label{sub:user_story}

In the span of two months, from October to November 2021, a fraudulent \ac{nft} creator and $\mathbb{X}$ user known as Homer\_eth executed five different \ac{nft} project rug pulls within two months, accumulating over \$2.8 million in profits.
Table~\ref{tab:homer_rug_pulls} provides an overview of Homer\_eth's rug pull projects, each with launch date and the estimated profit. 

\begin{table}[htb!]
  \centering
  \caption{Rug Pull Projects by Homer\_eth}
  \label{tab:homer_rug_pulls}
  \begin{tabular}{lcl}
    \toprule
    Project Name & Launch Date & Estimated Profit\\
    \midrule
    Ether Bananas & 10/07/2021 & \$125k\\
    Ether Monkeys & 10/11/2021 & \$1.77m\\
    Zombie Monkeys & 10/15/2021 & \$413k\\
    Ether Reapers & 10/20/2021 & \$282k\\
    \ac{eth} Banana Chips & 11/23/2021 & \$208k \\
  \bottomrule
\end{tabular}
\end{table}

The basis of this user story is the transaction graph depicted in Fig.~\ref{fig:use_case_tg} that provides a simplified visualization of the transaction flow across multiple NFT projects linked to Homer\_eth.
%
%The graph in Fig.~\ref{fig:use_case_tg} 
It highlights key components, including EOA Nodes (Externally Owned Accounts), which represent the multiple wallet addresses of the rug puller, and Deployer Nodes (Smart Contract Creators), with the 0xc8a6 address being the deployer for multiple fraudulent contracts. 
The links between addresses are established through various transaction, such as mint transactions (e.g., mintReaper, mintBananaChips), which indicate purchases; fund transfers (e.g., Transfer 65.61 ETH to 0xc8a6), showing proceeds flowing to personal wallets or exchanges; and contract deployments (e.g., Deploy Ether Reapers). The transaction graph makes a critical indication of fraudulent intent visible.
Instead of using a multisig treasury or project contract, funds were immediately funneled to a single address controlled by the rug puller.

Bob's story begins with a common enthusiasm for the burgeoning world of \acp{nft}. 
His journey into the \ac{nft} market is marked by excitement and optimism, spurred by the success stories he sees online. 
Homer\_eth, an \ac{nft} creator and $\mathbb{X}$ user, has caught the attention of many like Bob by sharing his \ac{nft} projects on $\mathbb{X}$. 
His first \ac{nft} collection was \textit{Ether Bananas}, consisting of 750 \acp{nft}, was launched on October 7, 2021.
Only four days later, on October 11, Homer\_eth continued with the release of \textit{Ether Monkeys}, followed by the release of \textit{Zombie Monkeys}.
The buzz around Homer\_eth's projects, especially \textit{Ether Monkeys}, which promised additional utility through a casino to gamble and a decentralized autonomous organization to govern the \acp{nft}, according to \cite{zachxbt2022}, draws Bob into the fray. 
Being relatively new to the \ac{nft} market, Bob views this as an opportunity not to be missed.
Bob bought his first \ac{nft} from Homer\_eth, an \textit{Ether Reapers}, and with that purchase, he was no longer just a bystander; he was now an active participant in Homer\_eth's growing community.

% FORCED POSITION
\begin{figure*}[b!]
\centerline{\includegraphics[width=\textwidth]{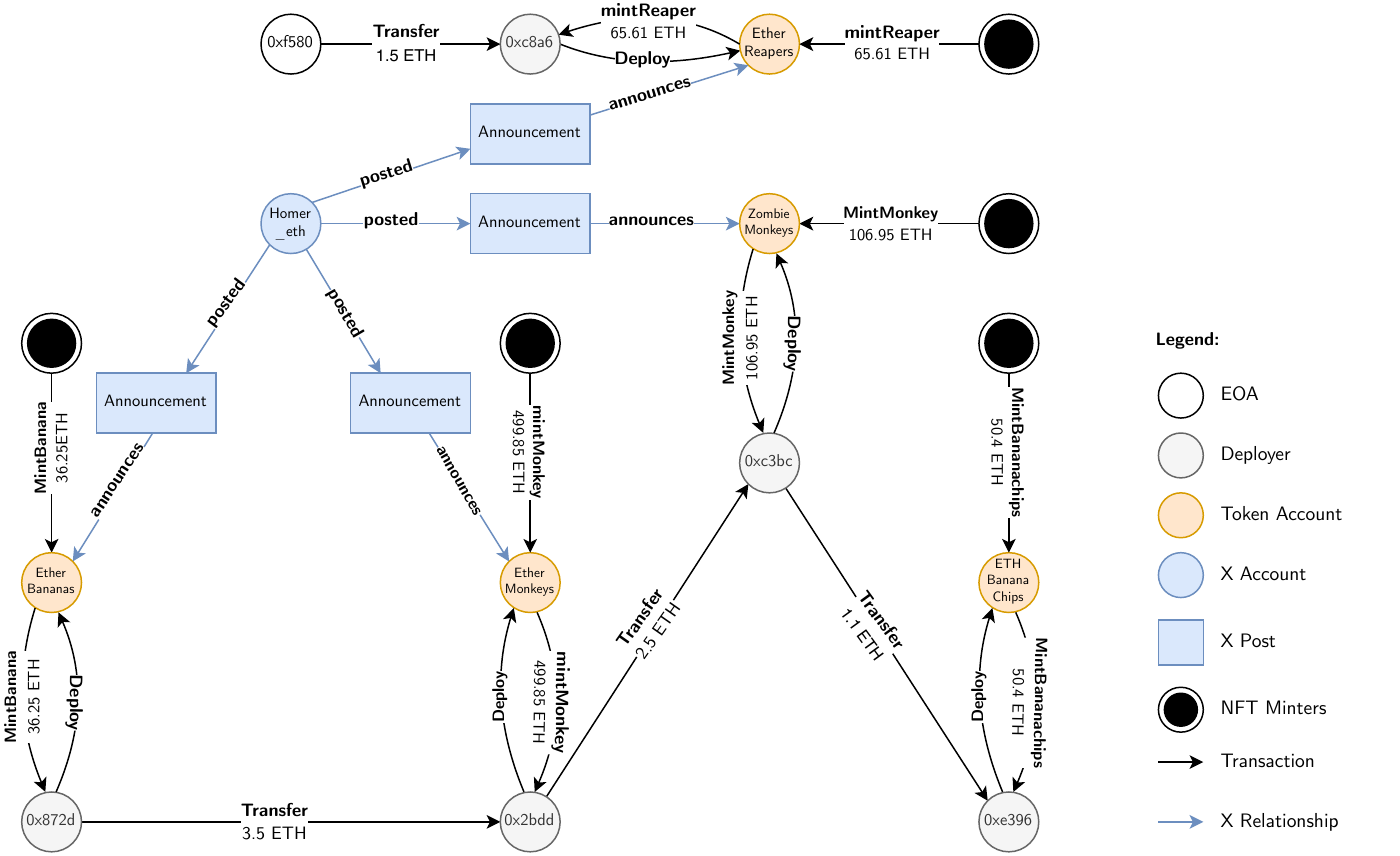}}
\caption{Knowledge graph of Homer\_eth's \ac{nft} rug pulls, constructed using Kosmosis.}
\label{fig:use_case_kg}
\end{figure*}

Bob's involvement in the community deepened over time. He engaged in discussions, shared his excitement with fellow members, and reveled in the rumors of more \ac{nft} launches in the future. His commitment paid off when he earned himself a whitelist spot that allows Bob to mint the upcoming \ac{nft} project \textit{\ac{eth} Banana Chips} by Homer\_eth. Convinced of its potential, Bob didn't hesitate to mint an \textit{\ac{eth} Banana Chips} \ac{nft} when the opportunity arose. With a click to confirm the transaction in his browser wallet (e.\,g., MetaMask \cite{metamask2023}), Bob became the proud owner of an \textit{\ac{eth} Banana Chips} \ac{nft}, unaware of the underlying risks associated with his investment.

However, the reality of the situation was far from the optimistic scenario Bob had envisioned. Unknown to him, since Bob had a limited understanding of blockchain transactions, the proceeds from the \textit{Ether Reapers} mint were not being locked in the smart contract for future development as promised. Instead, they were directly funneled into Homer\_eth's deployer address. From there, Homer\_eth will later transfer those mint proceeds either to his next deployer address for a future rug pull or to an exchange in order to pay out his profits made from rug pulling the projects.

After the launch of \textit{\ac{eth} Banana Chips}, a tense silence enveloped the community. For months, there was no news from Homer\_eth, no updates on the project, leaving everyone to wonder about the future. It wasn't until March 2022, that Homer\_eth broke the silence with the announcement of one last \ac{nft} project, dubbed \textit{Froggy Frens}. However, due to backlash from the community, Homer\_eth deleted his $\mathbb{X}$ account and vanished \cite{zachxbt2022}.

\subsection{Kosmosis Extension}

The basis of the extended user story is the Kosmosis \ac{kg}, depicted in Figure~\ref{fig:use_case_kg}.
Kosmosis identifies potential rug pulls by semantically analyzing transaction patterns encoded within smart contract interactions and cross-referencing blockchain addresses with real-world entity data from social media and other external sources. 
Our approach is grounded in the assumption that scammers publicly disclose or explicitly link blockchain addresses in their social media posts to promote their scams. 
This linkage is crucial for Kosmosis, as it provides the primary method of associating blockchain transactions with social identities, which enhances the semantic richness of the constructed \ac{kg}.
The Kosmosis \ac{kg} for this specific user story in Figure~\ref{fig:use_case_kg} is a direct enhancement over the basic transaction graph from Figure~\ref{fig:use_case_tg} as it was discussed in the previous subsection.
The enhancement comprises semantically annotated edges and the incorporation of data from the social media platform $\mathbb{X}$.

In order to detect potentially suspicious activity, we construct a Kosmosis subgraph using a multi-part SPARQL query. Listing~\ref{lst:construct} provides the overall CONSTRUCT query statement, capturing Ethereum transfers, mint transactions, smart contract deployments, bridging activities between chains, and social media references to blockchain addresses. 
Optional name labels help to identify the connection between blockchain or social media accounts and their associated real-world entities.

\begin{figure}[htbp]
\begin{lstlisting}[caption=Construction of the RDF subgraph., label=lst:construct]
PREFIX rdf:  <http://www.w3.org/1999/02/22-rdf-syntax-ns#>
PREFIX rdfs: <http://www.w3.org/2000/01/rdf-schema#>
PREFIX xsd:  <http://www.w3.org/2001/XMLSchema#>
PREFIX kos:  <http://oth-aw.de/kosmosis#>

CONSTRUCT {
  # transfers
  ?sender    kos:transferTo ?receiver .
  ?trEdge  a kos:TransferEdge ;
             kos:from  ?sender ;
             kos:to    ?receiver ;
             kos:value ?sumValueETH .

  # contract deployments
  ?deployer  kos:deployed ?contract .

  # mint flows 
  ?minter     kos:mint ?nftContract .
  ?mintEdge a kos:MintEdge ;
              kos:from  ?minter ;
              kos:to    ?nftContract ;
              kos:value ?sumMintETH .

  # UTXO unlock 
  ?utxoIn      kos:unlocks ?contractUtxo .

  # bridging
  ?bridgeFrom  kos:depositToBridge ?bridge .
  ?bridge      kos:bridgeTransfer  ?bridgeTo .

  # social layer
  ?xAccount    kos:posted   ?xPost .
  ?xPost       kos:announces ?announcedContract .

  ?anyAccount  kos:accountName ?accName .
}
\end{lstlisting}
\end{figure}

To enhance the interpretability of raw blockchain transactions, we introduce semantic annotations in our knowledge graph. 
This process involves using the Application Binary Interface (ABI) of smart contract transactions. 
Transactions interacting with NFT minting contracts contain function calls embedded in input data. 
Using the ABI, we extract function names (e.g., mintMonkey, mintBananaChips) and parameters. 
This enables labeling edges as minting transactions rather than generic asset transfers. 
Transaction type classification is done by categorizing transfers into value transactions, such as \textit{mintMonkey} and \textit{Transfer}, and non-value transactions, like contract deployments denoted as \textit{Deploy}. 
These semantics allow describing (i.\,e., tagging) sender and receiver addresses as \ac{nft} minter (previously depositor) and deployer (previously \ac{eoa}).

As part of the subgraph construction, multiple Ethereum value transactions between EOAs are aggregated in Listing~\ref{lst:aggregated-transfers}. The query calculates the total ETH transferred from each sender to each receiver and assigns a unique semantic identifier to these aggregated transfers. This process reduces transactional complexity while preserving critical relationships for identifying significant value flows  indicative of suspicious activity.
\vspace{0.01cm}
\begin{figure}[htbp]
\begin{lstlisting}[caption=Aggregated ETH transfers between EOAs on Ethereum., label=lst:aggregated-transfers]
WHERE {
  { 
    SELECT ?sender ?receiver (SUM(?value) AS ?sumValueETH)
    WHERE {
      ?tx a kos:ValueTransaction ;
            kos:from    ?sender ;
            kos:to      ?receiver ;
            kos:value   ?value ;
            kos:minedOn kos:Ethereum .
      ?sender   a kos:ExternallyOwnedAccount .
      ?receiver a kos:ExternallyOwnedAccount .
    }
    GROUP BY ?sender ?receiver
    HAVING (?sumValueETH > 0)
  }
  BIND( IRI(CONCAT("urn:tx-agg:",
        SHA256(CONCAT(STR(?sender),STR(?receiver))))) AS ?trEdge )
\end{lstlisting}
\end{figure}

Due to the categorization of blockchain transactions by their semantic functions further important transaction types can be captured.
Specifically, Listing~\ref{lst:transactions} collects direct Ethereum smart contract deployments, aggregates Ethereum NFT mint transactions according to function names embedded in transaction data, and incorporates Bitcoin UTXO transactions. 
Each aggregated transaction type is semantically annotated to clarify the nature of the underlying blockchain interaction.

Cross-chain interactions through bridge protocols are captured explicitly in Listing~\ref{lst:bridge}. 
Ethereum deposits made into bridge smart contracts and their subsequent transfers to Polygon-based externally owned accounts are identified and annotated. This detailed semantic labeling assists in tracing asset movements across blockchains, crucial for identifying potentially risky bridging behavior.
%Two-hop flow: Ethereum EOA deposits to bridge contract, bridge forwards to a Polygon EOA (purple node)

\begin{figure}[htbp]
\begin{lstlisting}[caption={Contract deployments and aggregated ‘mint’ calls on Ethereum, as well as UTXO transactions on Bitcoin.}, label=lst:transactions]
  # Direct deployments
  UNION
  {
    ?deployer bco:deployed ?contract .
  }

  # Mint transactions
  UNION
  { 
    SELECT ?minter ?nftContract (SUM(?val) AS ?sumMintETH)
    WHERE {
      ?mintTx a kos:CallTransaction ;
                kos:from  ?minter ;
                kos:calls ?nftContract ;
                kos:value ?val ;
                kos:action ?funcName .
      FILTER regex(?funcName, "^mint", "i")
      ?nftContract a kos:TokenAccount .
    }
    GROUP BY ?minter ?nftContract
  }
  BIND( IRI(CONCAT("urn:mint-agg:",
        SHA256(CONCAT(STR(?minter),STR(?nftContract))))) AS ?mintEdge )

  # UTXO transactions
  UNION
  {
    ?utxoIn a kos:TransactionInput ;
              kos:unlocks ?contractUtxo .
  }
\end{lstlisting}
\end{figure}

Finally, the data integrated from platform $\mathbb{X}$ enriches the \ac{kg} with detailed information about user accounts, labeled as \textit{X Account}, and specific announcements or posts, referred to as \textit{X Post}.
This integration facilitates a deeper understanding of the context and relationships surrounding these rug pulls. 
For instance, the \ac{kg} can establish a connection between previously unrelated entities, such as the deployer address \textit{0xc8a6} and the user Homer\_eth. 
This connection is made through a social media announcement in which Homer\_eth claims to have created the Ether Bananas project, as well as through semantic annotation, which identifies \textit{0xc8a6} as the deployer address of the Ether Bananas smart contract.

The final part of the query in Listing~\ref{lst:socials} retrieves social media data integrated from platform $\mathbb{X}$, focusing on posts that explicitly announce (e.g., direct claim of creation: ``Proud to announce the launch of Ether Bananas!'') or reference blockchain contracts in the comments section (e.g., ``The CA is 0xCeF4CCb03dbc7D87B388407e381a949bE6d00E3b,'' where CA stands for contract address of the NFT).  
It associates user identities ($\mathbb{X}$ accounts) and their posts with blockchain addresses they mention or promote. Optionally, the query includes usernames or labels from social accounts, further enhancing the contextualization of blockchain entities that allows to directly to real-world social identities, if possible.

\newpage

\begin{figure}[htbp]
\begin{lstlisting}[caption=Bridge deposits and transfers from and to the Ethereum blockchain., label=lst:bridge]
  UNION
  {
    ?depTx  a         kos:ValueTransaction ;
            kos:from  ?bridgeFrom ;
            kos:to    ?bridge ;
            kos:value ?depVal .
    ?bridge a kos:ContractAccount .

    ?transTx a         kos:ValueTransaction ;
             kos:from  ?bridge ;
             kos:to    ?bridgeTo ;
             kos:value ?depValPolygon .
    ?bridgeTo a kos:ExternallyOwnedAccount ;
              kos:existsOn kos:Polygon .
  }
\end{lstlisting}
\end{figure}

\begin{figure}[htbp]
\begin{lstlisting}[caption={$\mathbb{X}$ Posts that announce contracts. Optionally, carrying over any account name literals, if present.}, label=lst:socials]
  UNION
  {
    ?xAccount a kos:XAccount ;
                kos:posted ?xPost .
    ?xPost    a kos:XPost ;
                kos:announces ?announcedContract .
  }
  OPTIONAL { ?anyAccount bco:accountName ?accName }
}
\end{lstlisting}
\end{figure}

In conclusion, the Kosmosis \ac{kg} provides the semantic data foundation necessary for sophisticated detection logic. By aggregating transactions and enriching them with semantic annotations, the system can detect suspicious patterns, such as rapid bulk transfers following token minting events. Such transactions can be assigned elevated risk scores based on correlated indicators (e.g., rapid withdrawal to external accounts controlled by the deployer), triggering timely automated alerts and significantly reducing the reaction time required to prevent potential rug pulls.

\subsection{Future User Story} \label{sub:extended_user_story}

In an alternative scenario where Bob would have used Kosmosis, it would have analyzed the transaction history.
The system would have issued a rug pull warning based on patterns of fund diversion to deployer addresses.
Bob's journey in the \ac{nft} market would have been safer, beginning with his initial transaction to purchase an \textit{Ether Reapers} \ac{nft}. 

As soon as Bob initiated his transaction, the rug pull prevention mechanism would have accessed the \ac{kg}, to analyze the rug pull risk of the contract.
Based on the integrated knowledge from $\mathbb{X}$, the system would have been able to link the contract, Bob is about to interact with, to all of Homer\_eth’s prior blockchain activity.
The \ac{kg} would have revealed a critical anomaly. Instead of the mint proceeds being transferred to the contract address of the project for future development, they were being diverted to the \textit{Ether Reapers} deployer address via the \textit{MintReaper} function. 
With smart contracts acting as an automated and trustless intermediary, where the code of the contract dictates the flow of funds according to predefined rules, this pattern of fund diversion is absent in legitimate projects.
When funds are sent directly to a team member's address, in this case the deployer address, the funds can be moved to exchanges or other addresses with ease (i.\,e., pulling liquidity from the project without fulfilling the promises). 
This is a common tactic in rug pulls, where the developers abandon the project and disappear with the investor funds.
Therefore, signaling a potential rug pull behavior. 
Upon detecting this anomaly, the system would have immediately issued a rug pull warning to Bob, prompting Bob to make an informed decision by asking whether he wishes to proceed with the transaction despite the identified risk. 
This proactive approach empowers Bob to reconsider his decision with full awareness of the potential danger, offering him a chance to opt-out before potentially falling victim to a rug pull.

%\section{Future Work} \label{s:future_work}
\section{Current Limitations} \label{s:future_work}

We aim to translate the Kosmosis approach into a practical implementation.
The initial findings of our research on Kosmosis have shown promising results, indicating the potential of our approach in identifying and preventing rug pulls. 
However, there are ample improvement opportunities for Kosmosis in future work. 

The generalization, from the exemplary use case to a sophisticated general rug pull classification method, covering various data patterns in the \ac{kg}, is open research.
Our subsequent endeavor involves the development of an algorithm capable of discerning rug pull warnings at varying confidence levels.
This pursuit commences with the formulation of an intricate SPARQL query.
Furthermore, an alerting system that utilizes the \ac{kg}, constructed with Kosmosis, to alert users before interacting with a potential rug pull project, as described in the user story of Section~\ref{s:use_case}, requires future efforts.

It will be necessary to refine the filters used in the ingestion of data from the $\mathbb{X}$ Filtered Stream \ac{api}.
The current process of data ingestion depends on the presence of direct links to blockchain addresses in social media posts.
For instance, the ability to link the user Homer\_eth with the \textit{EtherReapers} smart contract was solely facilitated by the explicit mention of the smart contract address in Homer\_eth's announcement post on $\mathbb{X}$.
This example underscores the limitations of the current approach, which may overlook relevant connections in the absence of direct references. Consequently, a more sophisticated approach is required to ensure a broader and still relevant dataset is captured to associate $\mathbb{X}$ users with their respective blockchain addresses.

Additionally, the implementation of knowledge fusion, the process of identifying true subject-predicate-object triples \cite{dong2014}, sourced from the blockchain and social media stands out as a critical next step. 
By fusing multiple records representing the same \acl{rwe} into a single and consistent representation \cite{bleiholder2009}, knowledge fusion would allow for a more accurate representation of \aclp{rwe} in the knowledge graph.

Currently, our prototype is limited to blockchains utilizing the account-based accounting model, like Ethereum. 
Recognizing the diversity in blockchain architectures and their unique features, we aim to allow for the integration of blockchains using a different accounting system, like Bitcoin. 
This expansion is essential for broadening the applicability and utility of Kosmosis across different blockchain platforms.

\section{Conclusion and Future Work} \label{s:conclusion}

The Kosmosis approach represents a significant advancement in addressing the challenges associated with crypto rug pulls.
Our proposed approach offers enhanced capabilities for semantic analysis, allowing the identification of fraud patterns that traditional transaction graph methods cannot detect.

We outlined a user story, where a threat actor known as Homer\_eth executed five \ac{nft} project heists within two months, accumulating over \$2.8 million in profits.
In such a scenario, we showed that Kosmosis provides a knowledge graph that improves the detection of such fraudulent schemes carried out through sophisticated transaction patterns that might otherwise go unnoticed in related approaches, such as smart contract code analysis.
This capability helps users make informed decisions and avoid becoming victims of fraud.

We also demonstrated how to build a knowledge graph from blockchain and social media data using the Kosmosis approach to incremental knowledge graph construction.
Kosmosis becomes the basis for semantic querying and reasoning over a graph of entities and the relationships among them,
facilitating analyses for cybercrime and fraud prevention, with the current focus on rug pulls as a major fraud scheme.

Kosmosis pipeline supports the ingestion of unstructured, semi-structured, and structured data, as well as the ingestion of new data at different time intervals.
%It supports continuous ingestion in a stream-like fashion, incrementally, periodically, or event-based ingestion.
During construction, the semantics of blockchain transactions are extracted to address ``why'' and ``how'' crypto assets were transferred.
Thus, Kosmosis extends the traditional transaction graph into a \emph{semantically enhanced transaction graph} in which the sender and recipient are still pseudonyms.
By incrementally constructing a \emph{knowledge graph from blockchain and social media data}, Kosmosis also bridges the gap between pseudonymous transactions and real-world entities.

%\newpage
% ======== References =========
\begingroup
\sloppy
\printbibliography[notcategory=selfref]
\endgroup 

\end{document}